\documentclass[11pt]{article}
\usepackage{amsmath,amsfonts,amssymb,epsfig,color}
\begin{document}
\title{Geometric Complexity Theory III: on
deciding positivity of Littlewood-Richardson coefficients}
\author{
Dedicated to Sri Ramakrishna \\ \\
Ketan D. Mulmuley
 \\
The University of Chicago and I.I.T., Mumbai\footnote{Visiting faculty member}
 \\   \\ Milind Sohoni\\ I.I.T.,
Bombay}

\maketitle

\newtheorem{prop}{Proposition}%[section]
\newtheorem{claim}{Claim}
\newtheorem{goal}[prop]{Goal}
\newtheorem{theorem}[prop]{Theorem}
\newtheorem{hypo}[prop]{Hypothesis}
\newtheorem{problem}[prop]{Problem}
\newtheorem{question}[prop]{Question}
\newtheorem{remark}[prop]{Remark}
\newtheorem{lemma}[prop]{Lemma}
\newtheorem{cor}[prop]{Corollary}
\newtheorem{defn}[prop]{Definition}
\newtheorem{ex}[prop]{Example}
\newtheorem{conj}[prop]{Conjecture}
\newcommand{\ca}[1]{{\cal #1}}
\newcommand{\gt}[1]{{\langle  #1 |}}

\newcommand{\frcgc}[5]{\left(\begin{array}{ll} #1 &  \\ #2 & | #4 \\ #3 & | #5
\end{array}\right)}

\newcommand{\cgc}[6]{\left(\begin{array}{ll} #1 ;& \quad #3\\ #2 ; & \quad #4
\end{array}\right| \left. \begin{array}{l} #5 \\ #6 \end{array} \right)}

\newcommand{\wigner}[6]
{\left(\begin{array}{ll} #1 ;& \quad #3\\ #2 ; & \quad #4
\end{array}\right| \left. \begin{array}{l} #5 \\ #6 \end{array} \right)}

\newcommand{\rcgc}[9]{\left(\begin{array}{ll} #1 & \quad #4\\ #2  & \quad #5
\\ #3 &\quad #6
\end{array}\right| \left. \begin{array}{l} #7 \\ #8 \\#9 \end{array} \right)}

\newcommand{\srcgc}[4]{\left(\begin{array}{ll} #1 & \\ #2 & | #4  \\ #3 & |
\end{array}\right)}

\newcommand{\arr}[2]{\left(\begin{array}{l} #1 \\ #2   \end{array} \right)}
\newcommand{\ignore}[1]{}
\newcommand{\f}[2]{{\frac {#1} {#2}}}
\newcommand{\embed}[1]{{#1}^\phi}
\newcommand{\stab}{{\mbox {stab}}}
\newcommand{\perm}{{\mbox {perm}}}
\newcommand{\trace}{{\mbox {trace}}}
\newcommand{\polylog}{{\mbox {polylog}}}
\newcommand{\sign}{{\mbox {sign}}}
\newcommand{\poly}{{\mbox {poly}}}
\newcommand{\formula}{{\mbox {Formula}}}
\newcommand{\circuit}{{\mbox {Circuit}}}
\newcommand{\core}{{\mbox {core}}}
\newcommand{\orbit}{{\mbox {orbit}}}
\newcommand{\cycle}{{\mbox {cycle}}}
\newcommand{\ideal}{{\mbox {ideal}}}
\newcommand{\qed}{{\mbox {Q.E.D.}}}
\newcommand{\proof}{\noindent {\em Proof: }}
\newcommand{\weight}{{\mbox {wt}}}
\newcommand{\level}{{\mbox {level}}}
\newcommand{\vol}{{\mbox {vol}}}
\newcommand{\vect}{{\mbox {Vect}}}
\newcommand{\val}{{\mbox {wt}}}
\newcommand{\sym}{{\mbox {Sym}}}
\newcommand{\adm}{{\mbox {Adm}}}
\newcommand{\eval}{{\mbox {eval}}}
\newcommand{\for}{{\quad \mbox {for}\ }}
\newcommand{\mand}{{\quad \mbox {and}\ }}
\newcommand{\invlim}{{\mbox {lim}_\leftarrow}}
\newcommand{\directlim}{{\mbox {lim}_\rightarrow}}
\newcommand{\sformal}{{\cal S}^{\mbox f}}
\newcommand{\vformal}{{\cal V}^{\mbox f}}
\newcommand{\C}{\mathbb{C}}
\newcommand{\R}{\mathbb{R}}
\newcommand{\Q}{\mathbb{Q}}
\newcommand{\Z}{\mathbb{Z}}
\newcommand{\crystal}{\mbox{crystal}}
\newcommand{\conje}{\mbox{\bf Conj}}
\newcommand{\graph}{\mbox{graph}}
\newcommand{\rank}{\mbox{rank}}
\newcommand{\str}{\mbox{string}}
\newcommand{\RSK}{\mbox{RSK}}
\newcommand{\wt}{\mbox{wt}}

\begin{abstract} 
We point out that the remarkable Knutson and Tao Saturation Theorem 
 \cite{knutson} and polynomial time algorithms for linear 
programming \cite{lovasz}  have together an important, immediate consequence in 
 geometric complexity theory \cite{gct1,gcthyderabad}: The problem of
deciding positivity of Littlewood-Richardson coefficients belongs to $P$; cf.\cite{honey}.

Specifically, for $GL_n(\C)$,
 positivity of  a Littlewood-Richardson coefficient $c_{\alpha,\beta,\gamma}$
can be decided in time that is polynomial in $n$ and  the bit lengths of the specifications of 
the partitions $\alpha,\beta$ and $\gamma$. Furthermore, the algorithm is strongly 
polynomial in the sense of \cite{lovasz}. 

The main goal of this article is to explain the significance of this result in the context of
 geometric
complexity theory. Furthermore, 
it is also conjectured that an analogous result holds for arbitrary symmetrizable Kac-Moody 
algebras.
\end{abstract}

The fundamental Littlewood-Richardson rule in the representation theory of $GL_n(\C)$ 
\cite{fultonrepr}
states that the tensor product of two irreducible representations (Weyl modules) 
$V_{\alpha}$ and $V_{\beta}$ of $GL_n(\C)$
 decomposes as follows: 
\begin{equation} \label{eqdecomp1}
 V_{\alpha} \otimes V_{\beta} = \oplus_{\gamma} c_{\alpha,\beta,\gamma} V_{\gamma},
\end{equation}
where $c_{\alpha,\beta,\gamma}$ are Littlewood-Richardson coefficients.
Here $\alpha,\beta$ are partitions (Young diagrams) 
 with at most $n$ rows. The sum is over all Young diagrams   $\gamma$  of height at most $n$, and
 size  equal to the sum of the sizes of $\alpha$ and $\beta$.

This rule has been  studied intensively in representation theory; cf.
Fulton \cite{fultonyoung,fultonrepr}. But
the problem of deciding positivity 
of  $c_{\alpha,\beta,\gamma}$ 
efficiently did not receive much attention, perhaps because there was really no motivation
for studying  it.
The problem arises naturally in geometric complexity theory \cite{gct1,gcthyderabad,gct2},
 which is an approach to 
the fundamental problems in complexity theory (GCT), such as $P$ vs. $NP$, through algebraic 
geometry and representation theory. The basic philosophy of this approach is the 
{\em  flip} from  hard nonexistence to easy existence. Specifically,
the approach  first reduces the 
hard nonexistence  problems in complexity theory,  such as $P$ vs. $NP$, in characteristic 
zero, to showing existence of certain {\em obstructions}, which are certain 
gadgets with algebro-geometric and representation theoretic properties. The central 
geometric invariant theoretic \cite{git} results of GCT \cite{gcthyderabad,gct2} pave the road 
for proving easiness of this and related existence problems in
geometric invariant theory, once certain existence problems in representation theory
are shown to be easy. The transition from nonexistence to existence was proposed in
\cite{gct1}. The stronger transition from hard nonexistence to easy existence was 
proposed in \cite{gcthyderabad}, which is an extended abstract of \cite{gct2}. 

By divine justice, as was to be expected for the $P$ vs. $NP$ problem, showing that
these representation theoretic existence problems are easy turned out to be 
extremely hard. Because they are intimately related to the century old, fundamental 
unsolved problems of representation theory, such as the plethysm problem 
\cite{stanley,fultonrepr}.  As such, when the flip philosophy was first proposed in
\cite{gct1,gcthyderabad}, it went against the common belief among mathematicians.
Deciding positivity of a Littlewood-Richardson coefficient $c_{\alpha,\beta}^{\gamma}$--i.e.
 deciding if the 
the Weyl module $V_{\gamma}$ occurs (exists) within $V_{\alpha}\otimes V_{\beta}$--is the
simplest instance of  a general existence problem, called 
{\em  the subgroup restriction problem} in \cite{gcthyderabad} described below.
Its membership in $P$ (Theorem~\ref{tmain}) provides
the first concrete evidence in support of the flip philosophy of GCT.

It is a direct consequence of the Saturation Theorem of Knutson and Tao \cite{knutson} 
and polynomial time algorithms for linear programming \cite{lovasz}.
After a preliminary version of this note was  written, it was communicated to us by Prof. Tao that
actually they had thought briefly about the polynomial time algorithm
question for positivity in the different context of the Honeycomb model \cite{honey},
 and asked Peter Shor about it. 
He basically gave the same response that is in this note.
See page 180-181 of \cite{honey};
though there is a slight error in that paper, in asserting that
the simplex method takes polynomial time. 
 Nevertheless, as we point out,
 for the LP that arises here even a strongly polynomial time algorithm exists.

The Saturation Theorem itself was proved in an entirely different context: 
as
 a step in the proof of  Horn's conjecture \cite{zelevinsky,fultonhorn},
which arose from  the work of H. Weyl in 1912 and I. M. Gelfand in 1940's. 
After several attempts, finally Klyachko \cite{fultonhorn} proved some remarkable
results in the study of stability criterion for toric vector bundles on the projective plane.
 Zelevinsky observed \cite{zelevinsky} that  Horn's conjecture would follow from these
results  if the
Saturation Conjecture were proved; as happened soon after in \cite{knutson}.
For the sake of a computer scientist not familiar with these developments,
we give a self-contained proof of Theorem~\ref{tmain} here, assuming only the statement of the 
Saturation Theorem.

Theorem~\ref{tmain} 
 was stated in \cite{gcthyderabad} as known, implicitly assuming integrality of 
the polytope $P$ defined below. We recently realized that  $P$ need not be integral, 
in view of \cite{loera}, which disproved a conjecture of Berenstein and Kirillov 
\cite{berenstein} that
  Gelfand-Tsetlin polytopes are integral. Fortunately, 
the Saturation Theorem, which had come to our attention just then,
provided a sufficient relaxation of integrality.

Let $\lambda=(\lambda_1,\cdots,\lambda_k)$, where $\lambda_1 \ge \lambda_2 \ge \cdots \lambda_k >0$, be a partition (Young diagram).
By its  bit length,
we mean the bit length of its specification, which is 
$\sum_i \log_2(\lambda_i)$. 
Observe that the dimension of the Weyl module $V_{\lambda}$ 
 can be exponential 
in $n,k$ and  the bit lengths of $\lambda_i$'s. Because  the dimension 
of $V_\lambda$ is the total number of
semistandard tableau of shape $\lambda$ with entries in $[1,n]$ \cite{fultonrepr}.

\begin{theorem} \label{tmain}
Given partitions $\alpha,\beta$ and $\gamma$,
deciding if $V_{\gamma}$ exists  within $V_{\alpha} \otimes V_{\beta}$--i.e. if 
$c_{\alpha,\beta,\gamma}$  is positive--can be done
 in polynomial time; i.e., in time that is polynomial in $n$ and the bit lengths of $\alpha,\beta$,
and $\gamma$\footnote{If we assume that a partition $\lambda$ is specified as
$(\lambda_1,\cdots,\lambda_n)$, with $\lambda_1\ge \cdots \ge \lambda_n$, where 
$\lambda_i=0$ for $i$ higher than the height of $\lambda$, then the term $n$ can be subsumed
in the bit length of the input.}
 Furthermore, the algorithm is strongly 
polynomial in the sense of \cite{lovasz}.
\end{theorem} 

This is remarkable, since
 the dimensions of 
$V_{\alpha}, V_{\beta}, V_{\gamma}$ can  be exponential in $n$ and the bit lengths of
$\alpha_i,\beta_j$ and $\gamma_k$'s. What the result says is that whether an exponential 
dimensional object $V_{\gamma}$ can be embedded in another exponential dimensional 
object $V_{\alpha}\otimes V_{\beta}$ can be decided in time that is  polynomial in $n$ and the 
bit lengths of just their labels $\alpha,\beta$ and $\gamma$.

Strong polynomiality  stated in the theorem
 means that \cite{lovasz}: (1) The number of arithmetic steps in the algorithm 
  is polynomial in $n$.
It does not depend on the bit lengths of $\alpha_i,\beta_j$, and $\gamma_k$'s.
(3) The bit length of every intermediate operand that arises in the algorithm is 
polynomial in the total bit length of $\alpha,\beta$ and $\gamma$.

\subsubsection*{Subgroup restriction problem} 
The  fundamental  problems and conjectures  in representation 
theory that arise in geometric complexity theory are instances of the following 
{\em subgroup restriction problem}.
Suppose $G$ is a {\em reductive} group over $\C$ \cite{git}. 
In complexity theory, we shall only be interested in {\em nice} reductive groups such as:
 $SL_n(\C)$, the classical simple groups, the group $\C^*$ of nonzero complex numbers,
 finite simple groups, and the groups obtained from these 
by standard groups theoretic constructions
 such as products, wreath products etc. Suppose $H\subseteq G$ is a nicely
embedded, nice subgroup of $G$. Two important examples of nice embeddings are: 
\begin{enumerate} 
\item  $H \rightarrow G=H\times H$ (diagonal map). In this case, the subgroup restriction
problem will reduce to decomposing the tensor product of two representations of 
$H$, together with the associated decision problem.
\item $GL(\C^n)\times GL(\C^n) \rightarrow GL(\C^n\otimes \C^n)$. In this case, the subgroup
restriction problem will become equivalent to finding a positive decomposition of the
tensor product of two irreducible representations (Specht modules) of the symmetric group
\cite{stanley,fultonrepr}--a fundamental, century old
 unsolved problem in the representation theory
of the symmetric groups--together with the associated decision problem.
\item  $U$ is a representation of $H$,
$G=GL(U)$, and $H \rightarrow G$ is the representation homomorphism. 
In this case, the subgroup
restriction problem will reduce to the  (generalized) plethysm problem 
\cite{fultonrepr,stanley}--a fundamental, century old unsolved problem in the
representation theory of the general linear group--together with the 
associated decision problem.
\end{enumerate}
Let $V=V_\alpha$ be a representation of $G$, where $\alpha$ is  a label that completely 
specifies $V$. For example, if $G=GL_n(\C)$, and $V_\alpha$ is its 
irreducible representation (Weyl module) then the  label $\alpha$ is the
 Young diagram. If $V=V_\beta \otimes V_\gamma$, where $V_\beta$ and $V_\gamma$ are
irreducible, then the label $\alpha$ is the composite  $\beta \otimes \gamma$, and so on.

Since $H$ is a subgroup of $G$, $V$ is also a representation of $H$.
The classical result of H. Weyl says that $V$ has an essentially  unique  decomposition 
as an $H$-module: 
\begin{equation} \label{eqdecomp2}
 V_\alpha=\oplus_\beta m(\beta) W_\beta, 
\end{equation} 
where $\beta$ is the label ranging  over irreducible representations of $H$, 
$W_\beta$ is the corresponding irreducible representation, and 
$m(\beta)$ is its multiplicity. 

The {\em subgroup restriction problem} is find an explicit efficient
positive decomposition rule
for (\ref{eqdecomp2}) akin to the Littlewood-Richardson rule for (\ref{eqdecomp1}).
The associated existence problem is: 
 given
labels $\alpha$ and $\beta$ of $H$ and $G$ respectively, does
$W_\beta$ occur within $V_\alpha$? That is, is $m(\beta)$ positive? 
 The goal is to show that
this problem  belongs to the complexity class $P$. Here by polynomial, 
we mean polynomial in the numeric  parameters associated with $G$ and $H$
 and the  bit lengths of the labels 
$\alpha,\beta$.  For example, the numeric parameter associated with $GL_n(\C)$ is
$n$, with the symmetric group $S_n$ is $n$, and if $G$ is built using products etc. then
they are the numeric parameters of the building blocks.

When $H=GL_n(\C) \rightarrow  GL_n(\C) \times GL_n(\C)$, 
the decomposition (\ref{eqdecomp2}) coincides with the tensor product decomposition 
(\ref{eqdecomp1}), and the decision problem is simply deciding positivity of 
a Littlewood-Richardson coefficient. Though the general problem is far harder 
than  the latter, it is qualitatively similar.
Hence, Theorem~\ref{tmain} supports the conjecture 
 that the general problem also belongs to $P$.

Once this representation theoretic existence problem is shown to be in $P$,
and a sufficiently {\em concrete} \cite{gcthyderabad,gct2} form of the 
decomposition (\ref{eqdecomp2}) is found, 
the central algebro-geometric results of GCT in \cite{gcthyderabad,gct2}
give a lead on the (harder) geometric invariant theoretic  existence problems in
GCT. The road ahead is undoubtedly long and arduous, but, at least, the journey has begun.

\subsection*{Proof} %\label{sproof}
The proof of Theorem~\ref{tmain}  follows easily from the following three results:
\begin{enumerate} 
\item Littlewood-Richardson rule: specifically,
a polyhedral interpretation of the Littlewood-Richardson coefficients.
The polytope  we use here is  more elementary than Berenstein-Zelevinsky 
polytope \cite{berenstein2} and the Hive polytope \cite{knutson}--the latter two have 
some stronger properties not used here.
\item  Saturation Theorem \cite{knutson}.
\item Polynomial time algorithm for  linear programming: e.g. the ellipsoid 
 or the interior point
method, and the related strongly polynomial time algorithm for combinatorial
linear programming due to Tardos \cite{lovasz}.
\end{enumerate} 

Let us  begin with a polyhedral interpretation; this should be well known.
Recall that the Littlewood-Richardson coefficient  $c_{\alpha,\beta}^{\gamma}$ has the following
 combinatorial interpretation \cite{fultonyoung} .

Let us say that 
a word $w=w_1\cdots w_r$ is a {\em reverse lattice word} if, when read backwards from the
end to any letter $w_s$, $s<r$,  the sequence 
$w_r \cdots w_s$  contains at least as many $1$'s as $2$'s, at least 
as many $2$'s as $3$'s, and so on for all positive integers. 
The row word $w(T)$ of a skew tableau $T$  is defined to be the word obtained by reading 
its entries from bottom to top, and left to right. 
A skew-tableau $T$ of shape $\gamma/\alpha$ is called a {\em Littlewood-Richardson skew tableau}
 if 
its row word $w(T)$ is a reverse lattice word. 

Then  $c_{\alpha,\beta}^{\gamma}$  is the number of 
Littlewood-Richardson skew tableaux of shape $\gamma/\alpha$ of content $\beta$. 

Let $r_{j}^i(T)$, $i\le n$, $j\le n$, denote the number of $j$'s in the $i$-th row of 
$T$. These are integers satisfying the constraints: 
\begin{enumerate} 
\item Nonnegativity: $r^i_j \ge 0$.
\item Shape constraints: For $i \le n$,
\[  \alpha_i + \sum_j r^i_j = \gamma_i. \]
\item Content constraints: For $j\le n$: 
\[ \sum_i r^i_j=\beta_j.\] 
\item Tableau constraints: No $k \le j$  occurs in the row $i+1$ of $T$ 
below a $j$ or a higher integer in the row $i$ of $T$:
\[ \alpha_{i+1}+\sum_{k \le j} r^{i+1}_k \le \alpha_i + \sum_{k' <  j} r_{k'}^i.\]
\item Reverse lattice word constraints: 
$r^i_j =0$ for $i<j$, and for $i\le n$, $1<j\le n$:
\[ \sum_{i'le i} r^{i'}_j \le \sum_{i' < i} r^{i'}_{j-1}.\]
\end{enumerate} 

Let $r$ denote the vector with the entries $r_{j}^i(T)$. 
These constraints can be written in the form of a linear program: 
\begin{equation}  \label{elinearP}
A r \le b,
\end{equation}
where the entries of $A$ are $0,1$ or $-1$, and the entries of $b$ are homogeneous,
integral, linear forms in $\alpha_i,\beta_j$, and $\gamma_k$'s. 
Thus $c_{\alpha,\beta}^{\gamma}$ is the number of integer points in the polytope $P$ 
determined by these constraints. 

\begin{claim} 
The polytope $P$ contains an integer point iff it is nonempty.
\end{claim} 
\proof One direction is trivial. 

Suppose $P$ is nonempty. 
Since $b$ is homogeneous in $\alpha,\beta$ and $\gamma$, it follows that, for any positive
integer $q$, $c_{q\alpha,q\beta}^{q\gamma}$ is the number of integer points in the
scaled polytope $qP$.
All vertices of $P$ have rational coefficients. Hence,
for some positive integer $q$, the scaled polytope $qP$ has an integer point. It follows 
that, for this $q$,
$c_{q\alpha,q\beta}^{q\gamma}$ is positive. Saturation Theorem \cite{knutson} says that,
in this case,
$c_{\alpha,\beta}^{\gamma}$ is positive. Hence, $P$ contains an integer point.
\qed

Whether $P$ is nonempty can be determined in polynomial time using 
either the ellipsoid or the interior point algorithm for linear programming. Since the linear 
program (\ref{elinearP}) is combinatorial \cite{lovasz}, this can also be
done in strongly polynomial time using Tardos' algorithm \cite{lovasz}. 
This proves Theorem~\ref{tmain}.

It is of interest to know if there is a purely combinatorial algorithm for this problem
that does not use linear programming; i.e., one similar to the max-flow or weighted matching
problems in combinatorial optimization. The polytopes that arise in these combinatorial 
optimization problems are unimodular and integral--i.e., their vertices are integral
 \cite{lovasz}.
 In contrast, $P$ need not be integral.
This is known for the Gelfand-Tsetlin
polytope that arises in the study of Kostka numbers\cite{loera}, and also for 
the Hive polytope \cite{king}.
Unlike the hive polytope \cite{knutson}, $P$
need not even have an integral vertex. It is reasonable to conjecture that
there is 
 a polynomial time algorithm that provides an {\em integral proof} of positivity 
of $c_{\alpha,\beta}^{\gamma}$, in the form of an integral point in $P$. The above 
algorithm, as also the one in \cite{honey},
 only provides a {\em rational proof}; i.e., a rational point in $P$.

There is a generalization of the Littlewood-Richardson rule for arbitrary classical
Lie algebras, and also for symmetrizable Kac-Moody algebras
 \cite{kashiwara,nakashima,littelmann}. It was erroneously
stated in \cite{gcthyderabad} 
as known that the problem of deciding
  positivity of  generalized Littlewood-Richardson coefficients also belongs to $P$. 
But now we
 conjecture that this is so. 

Specifically,
 let ${\cal G}$ be a symmetrizable generalized Kac-Moody algebra \cite{kashiwara}, 
with rank $r$, which is  the dimension of the corresponding 
symmetrizable generalized Cartan matrix.
Let $V_{\alpha},V_{\beta}$ be two irreducible integrable representations of ${\cal G}$ 
with highest weights $\alpha$ and $\beta$. Then it is known
that 
\[V_{\alpha} \otimes V_{\beta} = \oplus_{\gamma} c_{\alpha,\beta}^\gamma V_{\gamma},\] 
where $c_{\alpha,\beta}^\gamma$ are generalized Littlewood-Richardson coefficients,
as defined in  \cite{kashiwara,nakashima,littelmann}.

\begin{conj} \label{conj}
Given fixed $\alpha,\beta,\gamma$, positivity of $c_{\alpha,\beta}^{\gamma}$ can
be decided in  polynomial  time; i.e. in time that is polynomial in the rank $r$ and
the bit lengths of the specifications of $\alpha,\beta,\gamma$. Furthermore, 
there exists a strongly polynomial time algorithm for the same.
\end{conj}

The proof here 
does not generalize, since the saturation conjecture is known to be false for the type $B,C,D$
\cite{elashvili,loera}. 
Hari Narayanan pointed out to us that  J. De Loera and T. McAllister \cite{loera}
 have recently made
some conjectures for  the hive polytopes for types $B, C,D$.
These may provide a step towards the proof of Conjecture~\ref{conj} for types 
$B,C,D$.

Given the fundamental importance of the $P$ vs. $NP$ question,
 we hope this note will
bring to a computer scientist's attention, similar, but far more formidable
conjectures of geometric complexity theory \cite{gcthyderabad}.
 The first few  instances of these
conjectures  say that  the
decision versions of the well known representation theoretic positivity problems \cite{stanley},
such as the plethysm problem, 
have (strongly) polynomial time algorithms. It also makes sense to make  similar conjectures
for other positivity problems in \cite{stanley} not considered in \cite{gcthyderabad},
such as the ones concerning Kazdan-Lusztig polynomials.

\end{document}